\def\ba{\begin{eqnarray}}
\def\ea{\end{eqnarray}}

\def\nn{\nonumber \\}

\documentclass[11pt,tightenlines,superscriptaddress,floatfixolumn,english,aps,notitlepage,nofootinbib]{revtex4-1} 
\usepackage[T1]{fontenc}
\usepackage[latin9]{inputenc}
\usepackage{amsmath}
\usepackage{amssymb}
\usepackage{graphicx}
\usepackage{esint}
\usepackage{color}
\usepackage{enumerate} 
\usepackage{hyperref} 

\makeatletter

\newcommand{\non}{\nonumber\\}
\newcommand{\gf}{\Gamma_F }
\newcommand{\gc}{\Gamma_c}

\makeatother

\begin{document}
\ifx\ave\undefined
\global\long\def\ave#1{\left\langle #1 \right\rangle }
\fi
\ifx\absol\undefined
\global\long\def\absol#1{\left| #1 \right|}
\fi
\ifx\mev\undefined
\global\long\def\mev{{\rm \, MeV}}
\fi
\ifx\gev\undefined
\global\long\def\gev{{\rm \, GeV}}
\fi
\ifx\dpp\undefined
\global\long\def\dpp#1{\frac{d^{3}#1}{(2\pi)^{3}}}
\fi
\ifx\mh\undefined
\global\long\def\mh{\hat{\mu}}
\fi
\ifx\dzb\undefined
\global\long\def\dzb{\Delta Z}
\fi
\ifx\dz\undefined
\global\long\def\dz{\Delta z}
\fi
\ifx\dtb\undefined
\global\long\def\dtb{\Delta T}
\fi
\ifx\dt\undefined
\global\long\def\dt{\Delta t}
\fi

\preprint{This line only printed with preprint option}

\title{Femtoscopy of stopped protons}

\author{Andrzej Bialas}
\email{bialas@th.if.uj.edu.pl}
\affiliation{M. Smoluchowski Institute of Physics,
Jagellonian University, Lojasiewicza 11,
30-348 Krak\'ow, Poland}

\author{Adam Bzdak}
\email{bzdak@fis.agh.edu.pl}
\affiliation{AGH University of Science and Technology, 
Faculty of Physics and Applied Computer Science,
30-059 Krak\'ow, Poland}

\author{Volker Koch}
\email{vkoch@lbl.gov}
\affiliation{Nuclear Science Division,
Lawrence Berkeley National Laboratory,
Berkeley, CA, 94720, USA}
\begin{abstract}

The longitudinal proton-proton  femtoscopy (HBT) correlation function, based on the idea that in a heavy ion collision at $\sqrt{s}\lesssim20\gev$ stopped protons are likely to be separated in configuration space, is evaluated. It shows a  characteristic oscillation which appears sufficiently pronounced to be accessible in experiment. The proposed  measurement  is essential for estimating the baryon density in the central rapidity region, and can be also viewed as an (almost) direct verification of the Lorentz contraction of the fast-moving nucleus.

\end{abstract}
\maketitle

\section{Introduction}

The search for a possible phase structure of QCD has been a focus
point in strong interaction research. Lattice QCD calculations have
established that for vanishing and small net-baryon density the transition
from hadrons to quarks and gluons is an analytic cross-over \cite{Aoki:2006we}.
The situation at larger baryon density, on the other hand, is less
clear, since at present lattice methods cannot access this region 
because of the fermion sign problem. Here one has to rely on model calculations
and a large class of these models do indeed predict a first-order phase
coexistence region which ends in a critical point (see, e.g., Ref. \cite{Stephanov:2007fk}
for an overview). 

In order to explore the region of large net-baryon density experimentally
one studies heavy ion collisions at moderate beam energies $\sqrt{s}\lesssim20\gev$,
where a sufficient amount of the incoming nucleons are stopped at
mid-rapidity in order to achieve the necessary baryon density. Indeed,
since produced baryons always come as baryon -- anti-baryon pairs,
the only means of producing a finite net baryon density is by stopping
the nucleons of the colliding nuclei. Thus, in order to explore the
QCD phase diagram at large baryon density, the question of baryon
stopping is essential to understand. In fact, stopping the baryons 
is only a necessary condition. In addition to being at mid-rapidity
in momentum space they also need to overlap in configuration space.

The mechanism by which the incoming nucleons are stopped is indeed
a very interesting question
\cite{Busza:1983rj,Busza:1989px,Kharzeev:1996sq,Capella:1996th,
  Anishetty:1980zp,Li:2016wzh}. However,
independent of the specific mechanism, it seems rather unphysical that
the nucleons are stopped instantaneously. Instead, it will take time
and space for the nucleons to decelerate. Therefore, it is rather unlikely
that the stopped nucleons will end up at $z\simeq0$, i.e., at the
point of the collision of the two nuclei. Instead, one would expect
that the nucleons from the right-going nucleus will end up at positions
in configuration space with $z>0$ and the left-going ones at $z<0$
so that the stopped nucleons may actually be distributed bi-modally
in configuration space. This observation was recently pointed out
in Ref. \cite{Bialas:2016epd}. Based on a simple string model, Ref.~\cite{Bialas:2016epd}
found that for collision energies $\sqrt{s}\gtrsim10\gev$ the stopped
nucleons actually will not overlap significantly in configuration space.
Of course this observation was based on a rather simple model and
it would be much better if this observation could be verified or ruled
out in experiment. This is the purpose of this paper, where we propose
to measure {\it longitudinal} Hanbury Brown Twiss (HBT) type correlations 
(also known as femtoscopy \cite{Lisa:2005dd}) of the stopped protons, i.e. protons at $y_{cm}\approx 0$  with  transverse momentum not exceeding, say, 1 GeV.
Since femtoscopy does not a priori distinguish between stopped and
produced protons, it is important to choose a collision energy which
is small enough for proton production to be negligible but sufficiently
high so that  the deceleration length is large enough for
the stopped protons to be separated in configuration space. Thus an
energy of $\sqrt{s}\simeq20\gev$ appears to be a good choice since
at this energy the anti-proton to proton ratio is still
very small, $\bar{p}/p\simeq0.1$ \cite{Adamczyk:2017iwn,Anticic:2010mp}.

This paper is organized as follows: In the next section we present
and discuss the source function based on the same simple string model
used in Ref. \cite{Bialas:2016epd} (corrected, however, for the Fermi
motion inside a nucleus). Next we calculate the resulting
femtoscopy correlation function before we close with a discussion of
the various issues and limitations of this study.

\section{The source function}

The essential ingredient for femtoscopy is the underlying source of
the the emitted particles, protons in our case. The source is the
phase-space distribution of emission points, which are typically the
points of the last interaction of the protons before they fly to the
detector. Clearly a quantitative calculation of such a source function
would require a sophisticated simulation. However, we believe that
certain semi-quantitative aspects can be discussed without such a treatment,
and it is this approach we will take in the following. 

As already eluded to in the Introduction, once two nucleons collide
it is very unlikely or possibly even unphysical for them to come to
a stop right at the collision point. Instead they will only come to
a stop after a certain distance and time.  

The distance  $\Delta z$ and  time
$\Delta t$ between the collision and the final space-time point $(z,t)$ where and when a nucleon acquires its final rapidity $y$, depends on the mechanism of deceleration and thus on the model used for its description. It is in general a function of the initial and final rapidity, $Y_{i}$ and $y$ as
well as the typical transverse mass, $M_{\perp}$ the nucleon acquires
after the collision. 

For a given collision space-time point $\left(z_{c},t_{c}\right)$ in the center-of-mass 
frame of the two nucleons we thus have
\ba
z & =z_{c}\pm\Delta z\left(Y_{i},y,M_{\perp}\right),\nn
t & =t_{c}+\Delta t\left(Y_{i},y,M_{\perp}\right),  \label {ztf}
\ea
where the plus sign refers to the right-going particles and the minus
sign to the left-going ones.  

One sees from (\ref{ztf}) that, in order to construct the source 
needed for femtoscopy, we need a distribution of the collision points
in space and time and a model or theory which determines $\Delta z$ and~$\Delta t$.

\subsection{Distribution of collision points}

Let us start with the collision point distribution. Here we follow
Ref. \cite{Bialas:2016epd} and assume that the distribution of nucleons 
inside the target and projectile nuclei can be reasonably described by a Gaussian.
In this case the longitudinal ($z$-direction) and transverse components
of the collision point distribution factorise and subsequently we
will concentrate on the collision point distribution
in the z-direction, which, following Ref. \cite{Bialas:2016epd}, we assume to be proportional to the overlap of the
distribution of the nucleons in the left- and right- moving nuclei.

We thus have 
\ba
W_c(z_c,t_c)\sim
e^{-\gamma^2[z_c-\zeta_L(t_c)]^2/R_L^2}\,e^{-\gamma^2[z_c-\zeta_R(t_c)]^2/R_R^2}
\, \Theta(t_c),
\ea
where
\ba
\zeta_L(t)=-\zeta_R(t)=\zeta_0-Vt; \qquad \zeta_0\gg R_{L,R}/\gamma,
\ea
are the positions of the centers  of the nuclei at the time $t$.
$\zeta_0$ and $-\zeta_0$ are positions of the centres of left-moving
and right-moving nuclei at $t=0$ before the nuclei have any 
contact with each other. This implies $\zeta_0\gg R_{L,R}/\gamma$
and $t_c\geq 0$. Also, $\gamma=\cosh(Y_{cm})$ denotes the Lorentz
contraction factor for the incoming nuclei in the center-of-mass
frame, which we are working in. 

\subsection {Distribution of nucleon emission points $z$ and $t$}

Consider first the right-movers. For  the distribution of $z$ and $t$, we have
\ba
W_R(z,t)&=&\int dz_c dt_c W_c(z_c,t_c) \delta( z-z_c-\Delta z) \delta(t-t_c-\Delta t)\nn 
& \sim & e^{-\gamma^2[z-\Delta z-Z]^2/R_L^2}\,e^{-\gamma^2[z-\Delta z+Z]^2/R_R^2}\,\Theta( t -\Delta t) \label {right}
\ea
with $Z\equiv \zeta_0-V(t-\Delta t )$.

For left-movers, the formula differs by the sign of $\Delta z$:
\ba
W_L(z,t) \sim e^{-\gamma^2[z+\Delta z-Z]^2/R_L^2}\,e^{-\gamma^2[z+\Delta z+Z]^2/R_R^2}\,\Theta( t -\Delta t).
\ea

For  identical nuclei we have
\ba
W(z,t;P_i,P_f) &=& W_L(z,t)+W_R(z,t) \non
& \sim& \left( e^{-[z+ \Delta z]^2/\Gamma_c^2}+e^{-[z- \Delta
    z]^2/\Gamma_c^2}\right)e^{-Z^2/\Gamma_c^2}\,\Theta( t-\Delta t),
\ea
where $\zeta_0\gg R/\gamma$, $\Gamma_c^2=R^2/2\gamma^2$, and the dependence on the initial and final momenta is implicit via $\Delta z$ and $\Delta t$. 

What remains then is to determine $\Delta z$ and $\Delta t$.  For nucleons with small transverse velocities, the simplest model is that of linear energy loss, as for instance
used in the Lund model \cite{Lund_model} or the Bremsstrahlung model
\cite{Stodolsky:1971qf}.  Using the conditions\footnote{The second condition expresses the equation of motion with the force equal to $\sigma$. It is exact when the transverse velocity of the nucleon vanishes. We have verified that for the nucleons of transverse momenta not exceeding 1 GeV the corrections are negligible.} 
\ba
dE/dz=\sigma,\;\;\;\;dP/dt =\sigma,    \label{fund}
\ea
where $\sigma$  denotes the energy loss per unit length
or string tension
one obtains  \cite{Bialas:2016epd}
\begin{align}
\Delta z\left(P_{i},\,P_{f},\,\sigma\right) & =\frac{E_{i}-E_{f}}{\sigma}\label{eq:delta_z_general}; \qquad E_f=M_\perp \cosh y ,\\
\Delta t\left(P_{i},\,P_{f},\,\sigma\right) & =\frac{P_{i}-P_{f}}{\sigma}\label{eq:delta_t_general}; \qquad P_f=M_\perp \sinh y .
\end{align}
Here   $P_{i}=M\sinh(Y_{i})$, $E_{i}= \sqrt{M^{2}+P_{i}^{2}}$, 
and $P_{f}=M_{\perp}\sinh(y)$, $E_{f}= \sqrt{M_{\perp}^{2}+P_{f}^{2}}$
are  the initial and final longitudinal momenta and energies 

These equations determine $\Delta z$ and $\Delta t$ for initial and final longitudinal momenta, $P_{i}$ and $P_{f}$, transverse mass of the final proton, $M_\perp$, and for a given rate of energy loss  $\sigma$ (string tension). In reality the string tension is not a constant but may fluctuate
from collision to collision (e.g., depending on number of constituent quarks wounded in
a given collision).  Since it is unlikely, however, that a nucleon
with only one or two wounded quarks may fully stop, the sample of the
nucleons with the final rapidity $y\approx 0$ is expected to be
largely dominated by those with three wounded quarks. Thus we shall
ignore fluctuations due to the string tension and take $\sigma = 3
\sigma_0$= 3 GeV/fm. For the transverse mass we subsequently will
chose a value of $M_{\perp}=1.2 \gev$ (we verified that the results are not sensitive to the actual value of $M_\perp$).

\subsection {Fermi motion}

In the case of a nucleus-nucleus collision, the nucleons inside the
target and projectile nuclei experience Fermi motion. Consequently, the initial momentum of the colliding nucleons is  distributed around the nominal (mean) value of the nucleus-nucleus collision. 
This  broadens the emission source in the longitudinal
spatial direction, and thus affects the femtoscopy signal. We have  
\ba
W_F(z,t,P_i,P_f)= \int dP_i G_F(P_i- \langle P_i \rangle) W(z,t; P_i,P_{f} ), \label{wf}
\ea
where   $G_F(P_i - \langle P_i \rangle)$  is the distribution of the actual initial momentum $P_i$  of the nucleon  around the average $\langle P_i \rangle$. We shall take it in the form
\ba
G_F(P_i-\langle P_i \rangle) \sim e^{-[P_i-\langle P_i \rangle]^2/\gf^2}; \qquad
\Gamma_F= \gamma \sqrt{\frac{2}{5}} k_{F} \simeq \gamma \, 165 \mev ,
\ea
where $k_F$ is the Fermi momentum.\footnote{The value of $\Gamma_F$ follows from the demand that the 
distribution $G_F$ exhibits the same variance as the Fermi gas with
the Fermi momentum $k_F$. Also, we ignore the small effect of the binding energy of the nucleons and
  instead assume that we can treat the nucleons as free particles.} 
Note that due to the Lorentz boost  the width of the distribution $\gf$ scales with the Lorentz
factor $\gamma$, $\gf \sim \gamma$.  This increases substantially the width of this distribution in the energy region of interest.

\section {The HBT correlation function}

The femtoscopic longitudinal correlation function  we are seeking is given by \cite{Pratt:1984su,Lisa:2005dd}
\ba
C(\delta q_z;\delta q_0) - 1 =- \frac12 \frac{|\Phi(\delta q_z;\delta q_0)|^2}
{|\Phi(\delta q_z=0;\delta q_0=0)|^2} ,  \label{c}
\ea
where $\delta q_z$ is the difference of the longitudinal momenta of the two protons, 
$\delta q_0$ is the difference of their energies and
\ba
\Phi(\delta q_z,\delta q_0;P_i,P_f )=\int_{-\infty}^\infty  dz e^{i z\delta q_z}
\int_{\Delta t}^\infty  dt e^{ -i t\delta q_0} W(z,t;P_i, P_f)
\ea
is the Fourier transform of the density. 

Since we are working with Gaussians, the Fourier transforms are straightforward. We have 
\ba
\Phi(\delta q_z,\delta q_0;P_i,P_f)\sim \cos [\delta q_z \Delta z ]
e^{-(\delta q_z \gc)^2/4} e^{- i\delta q_0(\Delta t +\zeta_0/V)}e^{-(\delta q_0 \gc)^2/(4V^2)} , \label{phi}
\ea
where for the Fourier transform in $t$ it was essential to use the condition $\zeta_0\gg R/\gamma$ which allowed us to integrate over time from $-\infty$.

Thus, the final result for the correlation function, including Fermi motion, is 
\ba
C_F(\delta q_z;\delta q_0) - 1 =- \frac12 \frac{|\Phi_F(\delta q_z;\delta q_0)|^2}
{|\Phi_F(\delta q_z=0;\delta q_0=0)|^2} ,  \label{cf}
\ea
where
\ba
\Phi_{F}(\delta q_{z},\delta q_{0};P_{i},P_{f})= \int
dP_{i}G_{F}(P_{i}-\langle P_i \rangle)\Phi (\delta q_{z};\delta q_{0};P_{i},P_{f}) \nn
\sim  e^{-\frac{\Gamma _{c}^{2}+\Gamma _{F}^{2}/\sigma ^{2}}{4}\left(
\delta q_{z}^{2}+\delta q_{0}^{2}\right) }
 \left[ \cos \left( \delta q_{z}\Delta Z\right) \cosh \left( \delta
q_{0}\delta q_{z}\frac{\Gamma _{F}^{2}}{2\sigma ^{2}}\right) +i\sin \left( \delta q_{z}\Delta Z\right) \sinh \left( \delta
q_{0}\delta q_{z}\frac{\Gamma _{F}^{2}}{2\sigma ^{2}}\right) \right] , \label{PhiF}
\ea
where we have omitted a common phase which does not play any role in
the correlation function.
Here we have assumed that the momentum dependence of the shift $\Delta
z$, Eq.~\eqref{eq:delta_z_general}, may be approximated by
\begin{align}
  \Delta z = \frac{E_{i}-E_{f}}{\sigma} = \frac{E_{i}-\langle E_{i} \rangle}{\sigma} +
  \frac{\langle E_{i} \rangle-E_{f}}{\sigma} \simeq  \frac{P_{i}-\langle P_{i} \rangle}{\sigma} +
  \Delta Z;\;\;\;\;\Delta Z \equiv  (\langle E_{i} \rangle - E_{f})/ \sigma ,
  \label{eq:deltaZ_approx}
\end{align}
where $\langle E_{i} \rangle$ is the mean (nominal) energy of the
incident nuclei, and  $\Delta Z$ denotes the shift in space  for the mean
energy. We further assumed that the velocity of the
nuclei in the c.m. frame is close to unity, $V\simeq 1$.  
For the energies under
consideration, $\sqrt{s}> 10 \gev $, these  should be very good
approximations and indeed we find that the resulting correlation
function agrees with the full result within a few per mille.

\section{Results}

\begin{figure}[t]
\includegraphics[width=0.45\textwidth]{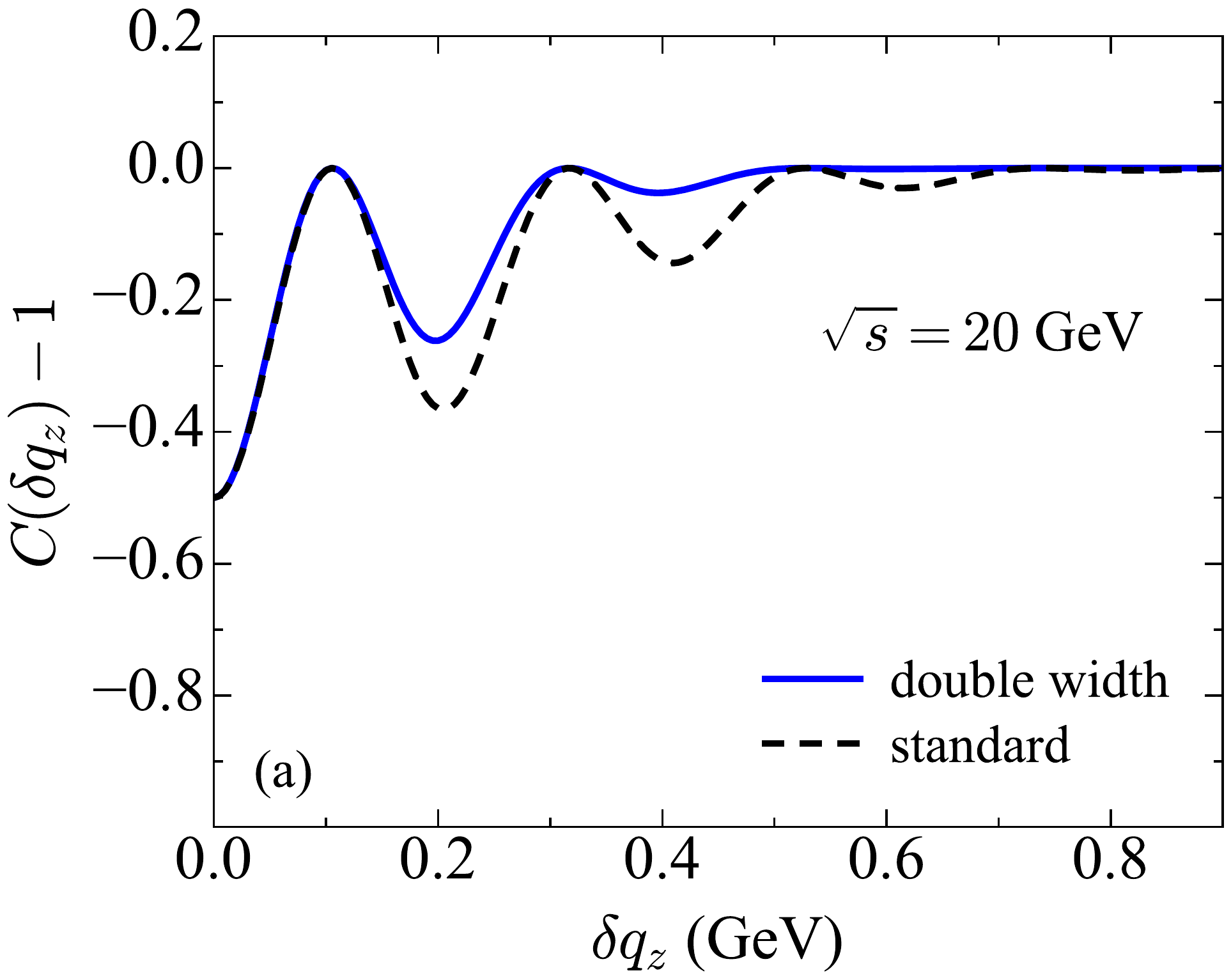}\hspace{6mm}
\includegraphics[width=0.45\textwidth]{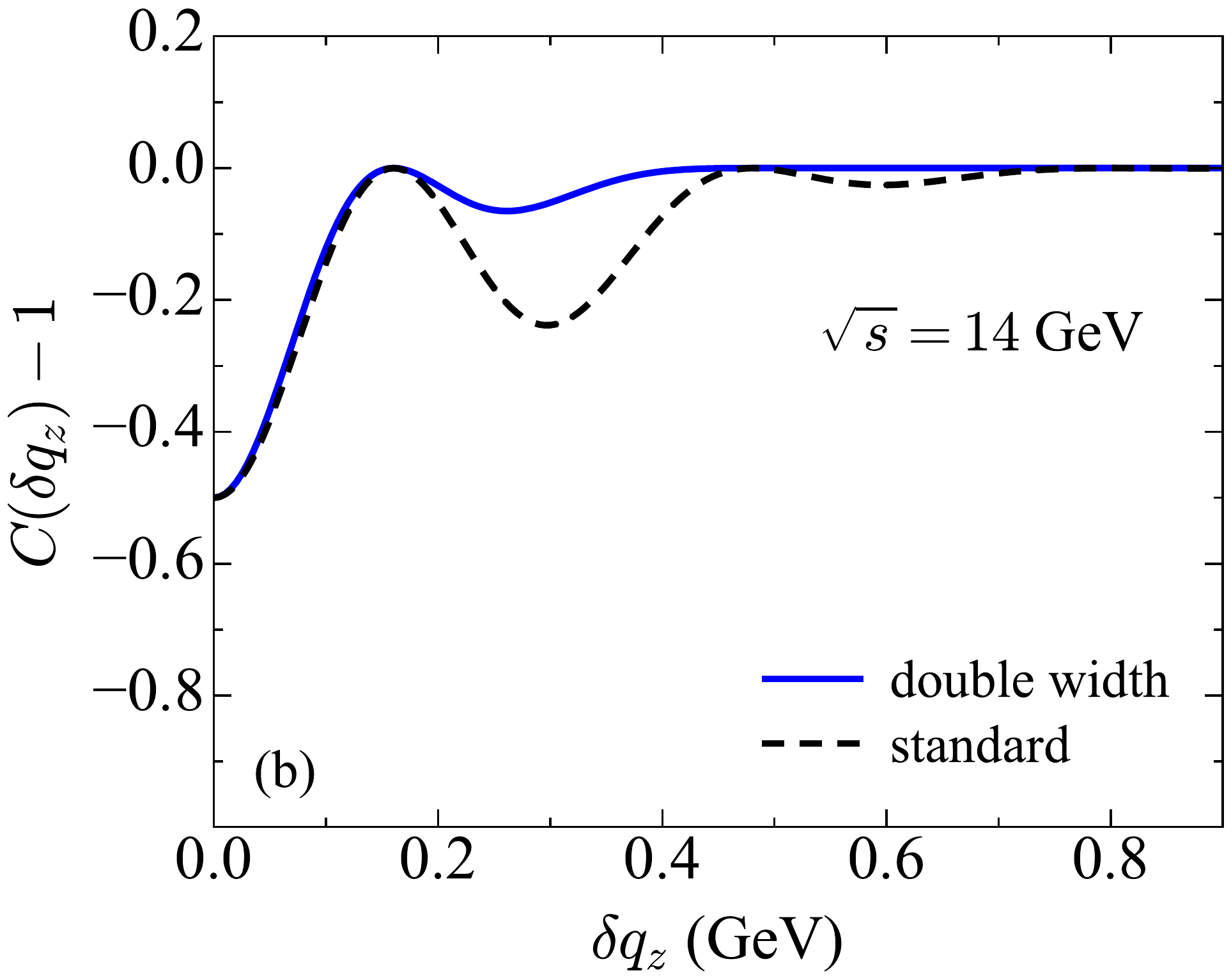}
\caption{Femtoscopy correlation function for (a) $\sqrt{s}=20\protect\gev$
and (b) $\sqrt{s}=14\protect\gev$. The black dashed lines represent the
result of our model calculation while the solid blue lines are obtained by
doubling the value of width of the collision point distribution, $\gc$.}
\label{fig:corr_result}

\hspace{10mm}

\includegraphics[width=0.45\textwidth]{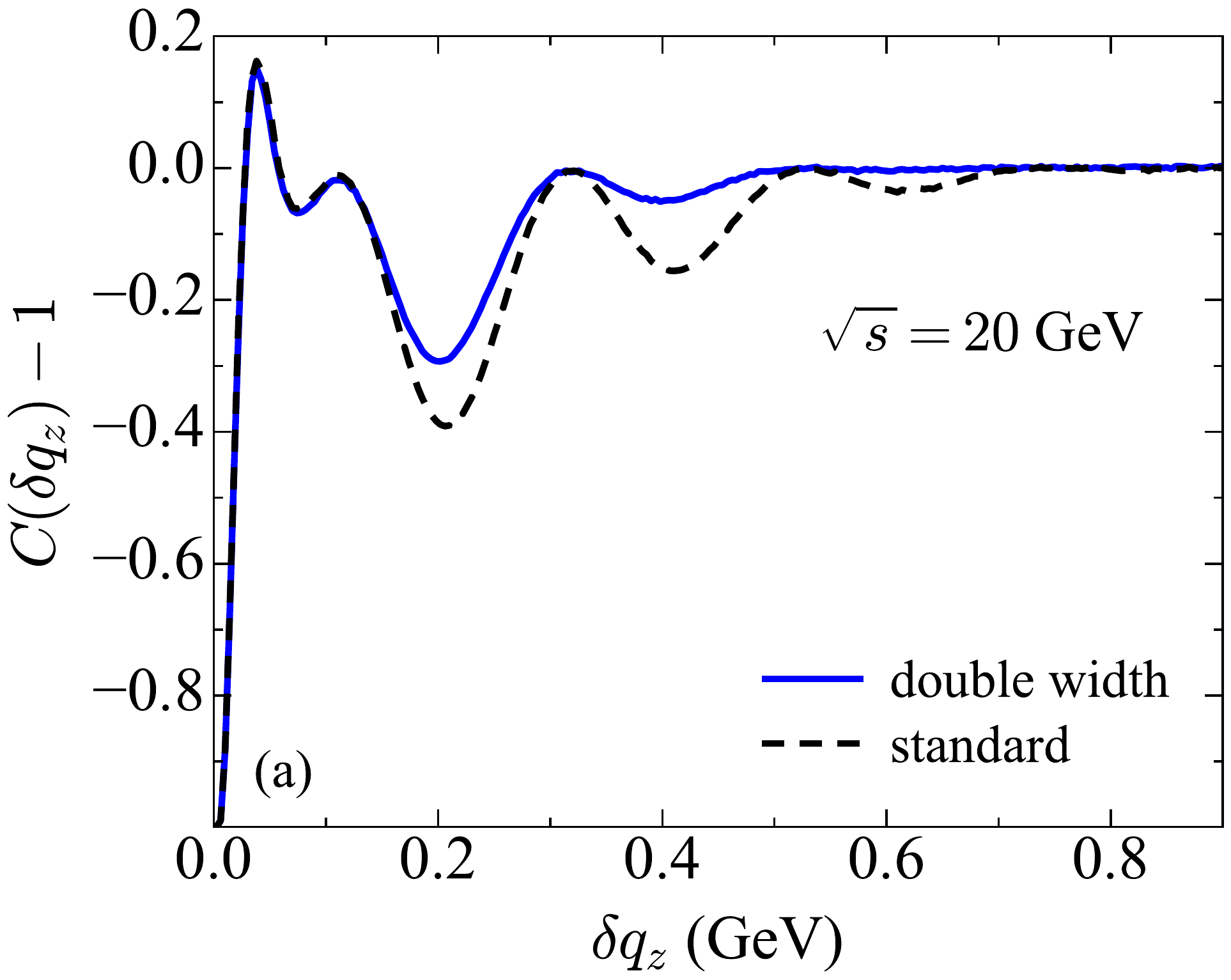}\hspace{6mm}
\includegraphics[width=0.45\textwidth]{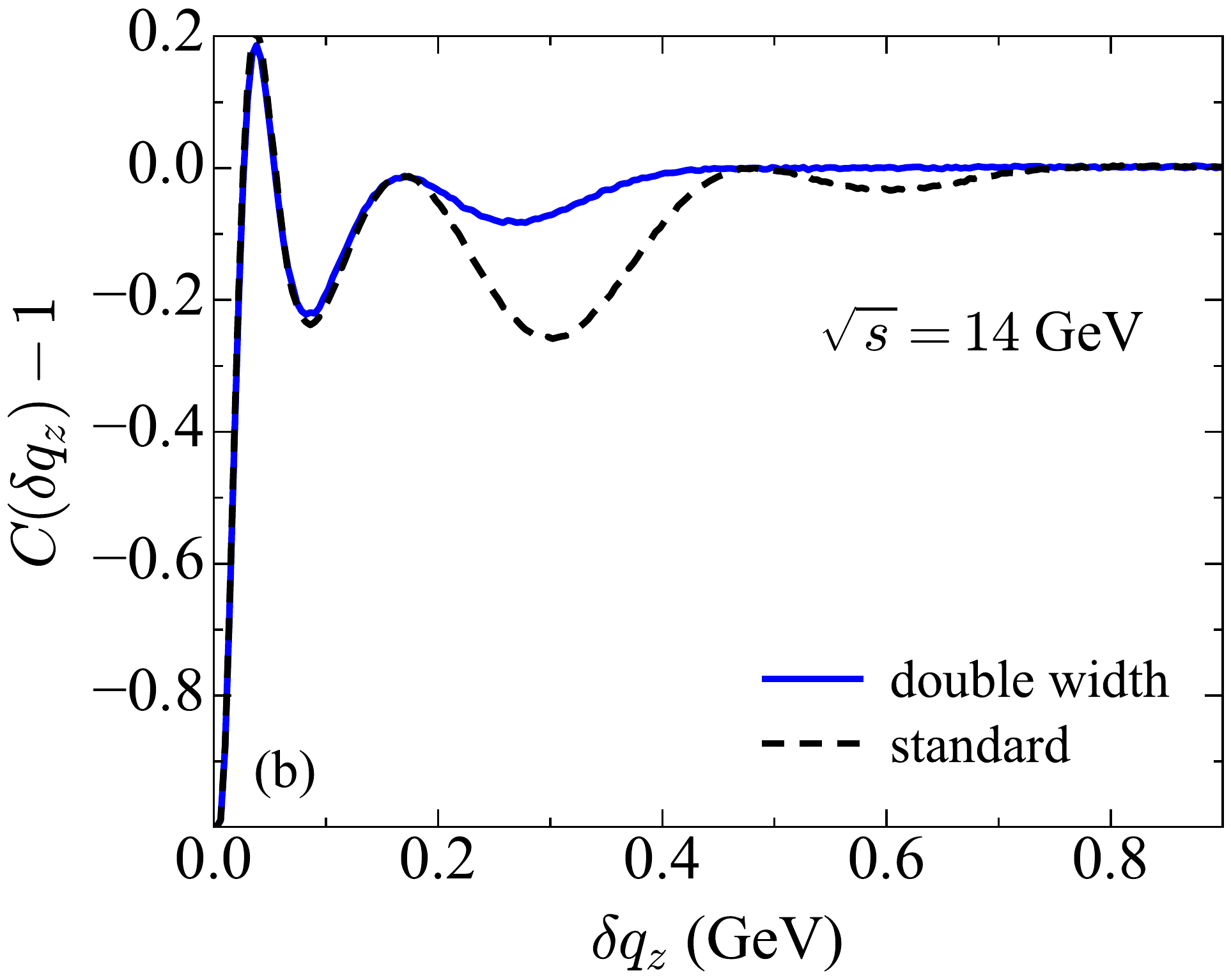}
\caption{Same as Fig. \ref{fig:corr_result} but with strong and Coulomb interaction effects included.}
\label{fig:corr_result_strong}
\end{figure}

The correlation function (\ref{cf}) evaluated for $\delta q_0=0$ (corresponding to protons with equal and opposite rapidities) is shown in
Fig. \ref{fig:corr_result}. 
Here we used for the radius $R=7 \, \mathrm{fm}$ and 
$M_{\perp,1}=M_{\perp,2} = 1.2$ GeV.
In panel (a) we show the predicted femtoscopy correlation function for
a collision energy of $\sqrt{s}=20\gev$ and in panel (b) for $\sqrt{s}=14\gev$.
The black dashed lines represent the result for the model discussed above.
The blue solid lines are the results, where we doubled the width $\Gamma_c$ of the
collision point distribution in order to allow for additional smearing (induced, e.g., by a non-zero proton radius not taken into account in our model). One sees characteristic oscillations of the correlation 
function, reflecting the two maximum structure of the source density.

To obtain experimental predictions, the correlation function seen in Fig. 1 must be corrected
for the final state interactions \cite{Lednicky:1981su,Adam:2015vja}. They are shown in Fig. 2 with strong and Coulomb interaction effects taken into account. One sees that although these corrections strongly affect the very small region of $\delta q_z$, the region of  $\delta q_z$ where the oscillations are observed remains qualitatively unchanged.

\begin{figure}[t]
\includegraphics[width=0.45\textwidth]{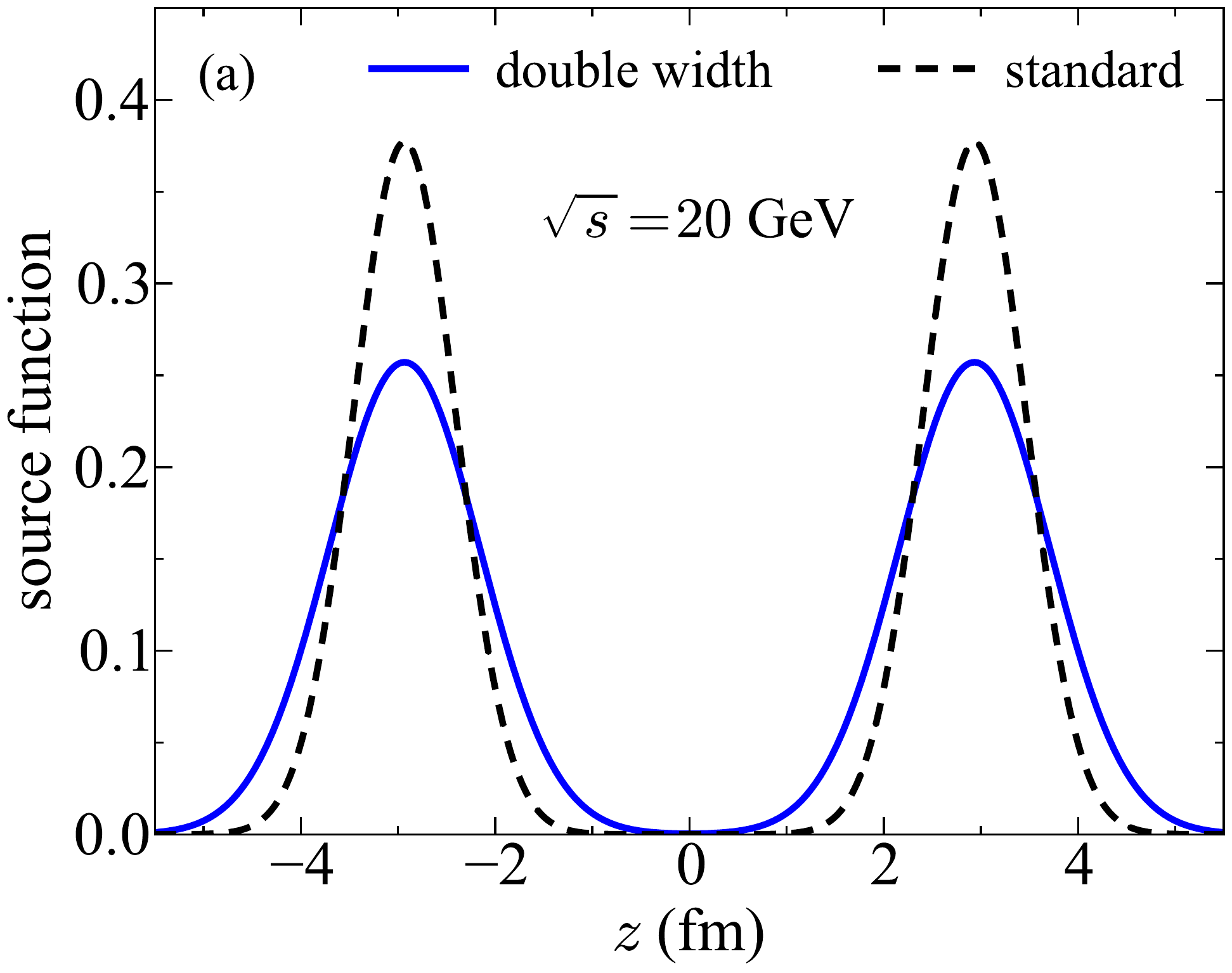}\hspace{6mm}
\includegraphics[width=0.45\textwidth]{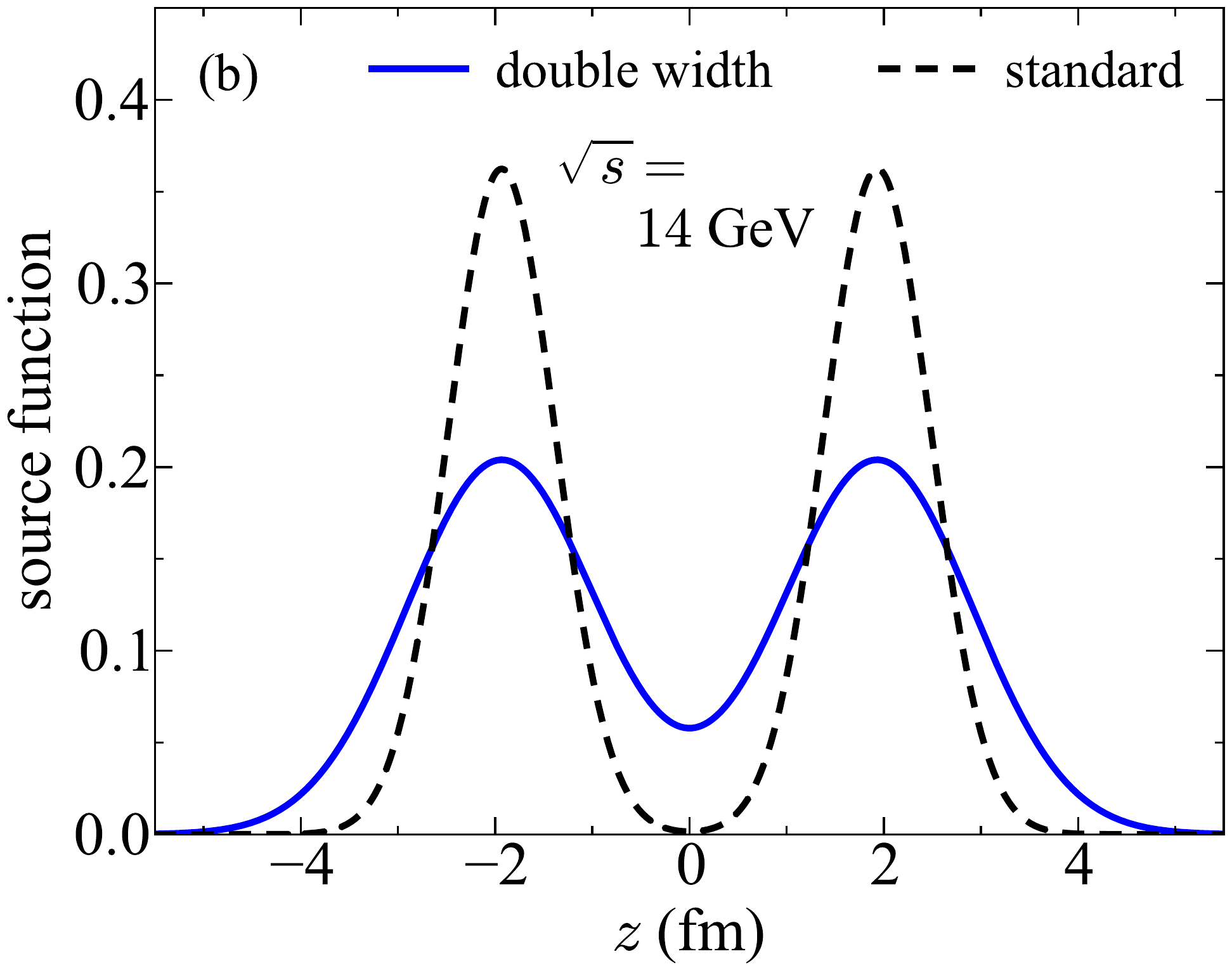}
\caption{Time integrated source function for stopped protons as a function
of $z$ for (a) $\sqrt{s}=20\protect\gev$ and (b) $\sqrt{s}=14\protect\gev$. 
The black dashed lines represent the result of our model calculation
while the blue solid lines are obtained by doubling the value of width of the
collision point distribution, $\gc$. The source functions shown are
normalized to unity.  
\label{fig:source_result}}
\end{figure}

In Fig. \ref{fig:source_result} we show the
corresponding time-integrated source distribution, 
\begin{align}
  \bar{W}(z;P_{i},P_{f}) &= \int dt  W_{F}(z,t; P_i,P_{f}) \non
  &\sim \exp\left( {-\frac{(z-\Delta Z)^{2}}{\gf^{2}/\sigma^{2} + \gc^{2}}} \right) 
    +  \exp\left(-{\frac{(z+\Delta Z)^{2}}{\gf^{2}/\sigma^{2} + \gc^{2}}}\right) .
  \label{eq:source_integrated}
\end{align}
Here we used the same approximation as before, Eq.~\eqref{eq:deltaZ_approx}.

We see that the separation of the stopped protons exhibited in the
source distribution manifests itself as extra oscillation in the
femtoscopy correlation function. For a collision energy of $\sqrt{s}=20\gev$
the signal is clearly visible for both the model result as well 
the more conservative result, where we doubled the width $\gc$. At 
$\sqrt{s}=14\gev$  the signal  is much weaker, however.

\section{Conclusion and Remarks}

In conclusion, we have presented a calculation of the longitudinal
femtoscopy correlation function of stopped protons based on the observation that in
a heavy ion collision at $10 \gev \lesssim \sqrt{s}\lesssim20\gev$ such protons are likely to be separated in configuration space.
The resulting correlation function shows extra oscillations which appear
sufficiently pronounced to be accessible in experiment. Clearly
such a measurement, if feasible, would be most desirable. It will
provide useful information about the longitudinal configuration space
distribution of the nucleons in a heavy ion collision, and, more importantly,
it will provide essential constraints on the mechanism by which baryon
number is transported to mid-rapidity. 

Some remarks are in order.
\begin{enumerate}[(i)]

\item  The observation of the  suggested extra oscillations will not only confirm the  idea  that the nucleons do not stop immediately after collision. It should  also allow to measure the effective distance at which the energy is deposited in the produced particles. Indeed, as seen from Eq. (\ref{PhiF}), $\Phi_F$ (and thus also $C_F$) explicitly depends on $\Delta Z$, the average distance required to stop a proton.

\item Even if  the oscillations are not seen, the measurement will
  determine the (longitudinal) size of the volume from which the
  protons at $y_{cm}\approx 0$ are emitted. This should allow to estimate the actual
  density of protons in configuration space, the quantity essential
  for the studies of this system.  One also obtains the upper limit on the  distance the nucleons travel before attaining  the rapidity $y\approx 0$, thus improving our understanding of the process of the energy loss by the leading particles in a high energy collision. 
 
\item  The definition of the longitudinal correlation function requires that the vector $\delta \vec{q}$ points in the $z$-direction, i.e. $\delta q_\perp = 0$. In our approximation  of the nuclear densities as Gaussians  this restriction is not important,  as the longitudinal and transverse degrees of freedom factorise.
To increase statistics, one may thus integrate over transverse momenta.
 Since the  Lund model is best justified at small transverse velocities, and since the Gaussian form is only an approximation, it seems   reasonable, however,   to restrict measurements to protons with  transverse momenta not exceeding, say, 1 GeV.  
  
\item It turns out that the corrections due to the Coulomb and strong interactions do not change
qualitatively the possibility of observation of the expected oscillations of the correlation
function. 

\item Our calculation ignored entirely possible correlations 
between the outgoing protons due to quark mixing at very short
distances \cite{bz}. Introducing such correlations may result in the correlation function being positive in some region of $\delta q_z$. As shown in Ref. \cite{bz}, however, this effect is small and should not modify our conclusions.

\item Finally, let us add that our results rely strongly on the idea that the longitudinal distribution of nucleons inside moving nucleus are Lorentz contracted and that this contraction survives during the collision. The proposed measurement should thus provide an interesting test of this effect (for the recent discussion of the measurements of Lorentz contraction, see Ref. \cite{Rafelski:2017yob}).

\end{enumerate}

\begin{acknowledgments}
We thank Scott Pratt for providing his code to calculate the effects of
strong and Coulomb interaction effects and useful correspondence.
Thanks are due to Mike Lisa for interesting discussions.
This work was partially supported by the
Faculty of Physics and Applied Computer Science AGH UST statutory tasks No. 11.11.220.01/1 within
subsidy of Ministry of Science and Higher Education, and by the National Science Centre, Grants
No. 2014/15/B/ST2/00175 and No. 2013/09/B/ST2/00497,
and the U.S. Department of Energy, Office of Science, Office of Nuclear Physics, 
under contract number DE-AC02-05CH11231.
\end{acknowledgments}

\end{document}